\documentclass[aip,jap,reprint,superscriptaddress,showpacs]{revtex4-1}

\usepackage{graphicx}
\usepackage{dcolumn}
\usepackage{bm}

\begin{document}

\title{Heat switch effect in an antiferromagnetic insulator Co$_3$V$_2$O$_8$}

\author{X. Zhao}
\affiliation{School of Physical Sciences, University of Science and Technology of China, Hefei, Anhui 230026, People's Republic of China}

\author{J. C. Wu}
\affiliation{Hefei National Laboratory for Physical Sciences at Microscale, University of Science and Technology of China, Hefei, Anhui 230026, People's Republic of China}

\author{Z. Y. Zhao}
\affiliation{Hefei National Laboratory for Physical Sciences at Microscale, University of Science and Technology of China, Hefei, Anhui 230026, People's Republic of China}

\author{Z. Z. He}
\affiliation{Fujian Institute of Research on the Structure of Matter, Chinese Academy of Sciences, Fuzhou, Fujian 350002, People's Republic of China}

\author{J. D. Song}
\affiliation{Hefei National Laboratory for Physical Sciences at Microscale, University of Science and Technology of China, Hefei, Anhui 230026, People's Republic of China}

\author{J. Y. Zhao}
\affiliation{Hefei National Laboratory for Physical Sciences at Microscale, University of Science and Technology of China, Hefei, Anhui 230026, People's Republic of China}

\author{X. G. Liu}
\email{lxgrz@ustc.edu.cn}

\affiliation{Hefei National Laboratory for Physical Sciences at Microscale, University of Science and Technology of China, Hefei, Anhui 230026, People's Republic of China}

\author{X. F. Sun}
\email{xfsun@ustc.edu.cn}

\affiliation{Hefei National Laboratory for Physical Sciences at Microscale, University of Science and Technology of China, Hefei, Anhui 230026, People's Republic of China}

\affiliation{Key Laboratory of Strongly-Coupled Quantum Matter Physics, Chinese Academy of Sciences, Hefei, Anhui 230026, People's Republic of China}

\affiliation{Collaborative Innovation Center of Advanced Microstructures, Nanjing, Jiangsu 210093, People's Republic of China}

\author{X. G. Li}

\affiliation{Hefei National Laboratory for Physical Sciences at Microscale, University of Science and Technology of China, Hefei, Anhui 230026, People's Republic of China}

\affiliation{Department of Physics, University of Science and Technology of China, Hefei, Anhui 230026, People's Republic of China}

\affiliation{Collaborative Innovation Center of Advanced Microstructures, Nanjing, Jiangsu 210093, People's Republic of China}

\date{\today}

\begin{abstract}

We report a heat switch effect in single crystals of an antiferromagnet Co$_3$V$_2$O$_8$, that is, the thermal conductivity ($\kappa$) can be changed with magnetic field in an extremely large scale. Due to successive magnetic phase transitions at 12--6 K, the zero-field $\kappa(T)$ displays a deep minimum at 6.7 K and rather small magnitude at low temperatures. Both the temperature and field dependencies of $\kappa$ demonstrate that the phonons are strongly scattered at the regime of magnetic phase transitions. Magnetic field can suppress magnetic scattering effect and significantly recover the phonon thermal conductivity. In particular, a 14 T field along the $a$ axis increases the $\kappa$ at 7.5 K up to 100 times. For $H \parallel c$, the magnitude of $\kappa$ can be suppressed down to $\sim$ 8\% at some field-induced transition and can be enhanced up to 20 times at 14 T. The present results demonstrate that it is possible to design a kind of heat switch in the family of magnetic materials.

\end{abstract}

\pacs{66.70.-f, 75.47.-m, 75.50.-y}

\maketitle

Heat switches are devices that switch as needed between roles of good thermal conductors and good thermal insulators. They have wide and important applications in not only the cryogenics but also the deep space detectors, space coolers and spacecrafts.\cite{DiPirro, Bartlett, Bosisio, Gopinadhan} A famous heat switch used in cryogenic cryostat is made by metal superconductors, whose electronic thermal conductivity can be switched on by applying magnetic field to suppress the superconductivity.\cite{DiPirro} This kind of heat switch works only at very low temperatures (well below the superconducting transition temperatures or at subKelvin temperatures). Since the phonon thermal conductivity at such low temperatures are negligibly small, the total thermal conductivity can be changed by several orders of magnitude with magnetic field, a nearly ideal thermal insulator to good thermal conductor transition. At not very low temperatures, the thermal conductivity ($\kappa$) of any materials usually cannot be significantly changed by control parameter,\cite{Jin, Hohensee} since there is always sizeable phonon thermal conductivity which is not easily changed much. However, past decades studies on the heat transport of magnetic materials have revealed that the phonon thermal conductivity can be remarkably changed by magnetic field in the case there is strong scattering between phonons and magnetic excitations.\cite{Ando, Berggold, Yan, Wang_HMO, Li_TTO, Gofryk} In particular, a large magnetic-field-induced increase of $\kappa$ can be observed in rare-earth manganites and titanates.\cite{Wang_HMO, Li_TTO} In this Letter, we report exceptionally large magnetothermal conductivity in an antiferromagnetic insulator, Co$_3$V$_2$O$_8$. As large as 100 times increase of $\kappa$ in magnetic field was found at temperature of several kelvins. The result provides a possible route to find heat switch devices that can work at not very low temperatures.

Co$_3$V$_2$O$_8$ has a crystal lattice with space group $Cmca$.\cite{Rogado, Sauerbrei} The edge sharing CoO$_6$ octahedra form a staircase kagom\'e structure and the kagom\'e staircase lattices are separated by the nonmagnetic VO$_4$ tetrahedra (see Fig. S1 of the Supplementary Material\cite{Supplementary}). The presence of two different Co$^{2+}$ sites along with competing interactions such as the single-ion anisotropy, the nearest-neighbor and the next nearest-neighbor exchanges, and Dzyaloshinskii-Moriya (DM) interactions lead to fascinating magnetic behaviors at low temperatures.\cite{Rogado, Sauerbrei, Szymczak, Chen, Wilson1, Yasui, Wilson2, Yen, Helton, Fritsch} In zero field, there are successive magnetic transitions from the paramagnetic (PM) to incommensurate antiferromagnetic (ICAF), commensurate antiferromagnetic (CAF) and commensurate ferromagnetic phases (CF).\cite{Yasui, Yen} The geometric frustration results in an antiferromagnetic order of Co$^{2+}$ spins at a rather low temperature of $T_N$ = 11.4 K.\cite{Rogado, Szymczak, Wilson1} At $T_N$, only the Co$^{2+}$ spins locating in the spine site order antiferromagnetically along the $a$ axis. The various phases at lower temperatures are distinguished by the commensurability of the spin density wave vector.\cite{Chen, Helton} The $b$-component of the magnetic modulation vector, $\delta$, decreases continuously from $\delta$ = 0.55 and it locks in at a commensurate value of $\delta$ = 1/2 at $T_{c1}$ = 8.9 K. Below $T_{c2}$ = 7.0 K, $\delta$ decreases continuously again locks into another commensurate value, $\delta$ = 1/3, at $T_{c3}$ = 6.6 K. At $T_{c4}$ = 6.2 K, $\delta$ becomes zero and the Co$^{2+}$ spins on both sites become ferromagnetically ordered with the spin alignment along the $a$ axis.

Since the Co$^{2+}$ spin has strong anisotropy, with the easy and hard axes along the $a$ and $b$ axis, respectively, the magnetic field along different crystallographical axes affects these phases in very different ways.\cite{Szymczak, Yasui, Wilson2, Yen, Fritsch} (See Fig. S1 of the Supplementary Material.\cite{Supplementary}) With $H \parallel a$, the ICAF and CAF phases disappear for $\mu_0H >$ 0.5 T and the CF phase extends to the higher temperature region, because the ground-state ferromagnetic structure is stabilized by the field. The magnetic states are insensitive to the field along the $b$ axis. The $H-T$ phase diagram with $H \parallel c$ was found to be most complicated, where the phases with $\delta$ = 1 and 2/3 exist in the region of $\mu_0H >$ 1 T. At the lowest temperature, the $c$-axis field drives phase transitions from the CF state to the ICAF state ($\delta$ = 1) and then to the PM (polarized) state at about 1 and 5 T, respectively.

Co$_3$V$_2$O$_8$ single crystals were grown by a flux method \cite{He}. The as-grown crystals are dark blue with size up to $3 \times 3 \times 1$ mm$^3$, with the thickness along the $b$ axis. For anisotropic $\kappa$ measurements, the long-bar shaped samples were cut from the as-grown crystals along either the $a$ or $c$ axis, respectively. The thermal conductivity was measured by using a conventional steady-state technique.\cite{Li_TTO, Wang_HMO, Chen_MCCL, Wang_TMO, Wu_CHC} In these measurements, the magnetic field is parallel to the heat current ($J \rm_H$), which is along the lengths of the samples. The specific heat was measured by the relaxation method using a commercial physical property measurement system (PPMS, Quantum Design). The magnetization was measured at low temperatures by using a SQUID-VSM (Quantum Design).

\begin{figure}
\includegraphics[clip,width=8.5cm]{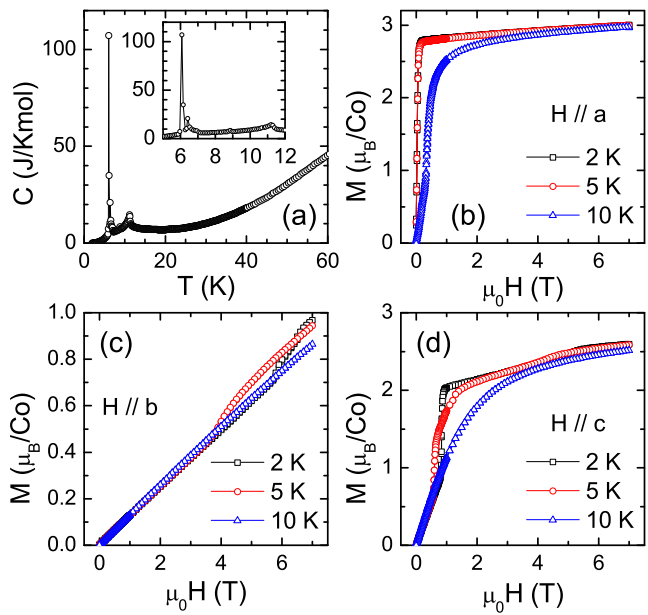}
\caption{(Color online) (a) Specific heat of Co$_3$V$_2$O$_8$ single crystal. Inset shows the zoom-in of data at 4--12 K, which shows five peaks or anomalies at 11.2, 8.8, 6.9, 6.4 and 6.0 K. (b--d) Magnetization curves of Co$_3$V$_2$O$_8$ single crystals at 2, 5 and 10 K and with magnetic field along three crystallographic directions. The data were taken with sweeping field 0 $\rightarrow$ 7 $\rightarrow$ 0 T and display no obvious hysteresis.}
\end{figure}

Our single crystals have been well characterized in an earlier work.\cite{He} Figure 1(a) shows low-temperature specific heat in magnetic field. Five peaks or anomalies can be seen at 11.2, 8.8, 6.9, 6.4 and 6.0 K, corresponding to five magnetic transitions. Figures 1(b--d) show the magnetization curves of our Co$_3$V$_2$O$_8$ single crystals at 2, 5 and 10 K for magnetic field along three crystallographic directions. For $H \parallel a$, the magnetization shows a typical behavior of a ferromagnetic phase. The value of the saturation magnetization is close to 3 $\mu_B$/Co, indicating a collinear ferromagnetic order along the $a$ axis. For $H \parallel b$, the magnetization is much smaller. At 10 K, the magnetization shows a nearly linear behavior, while some jumps appear at lower temperatures. At 5 K, a jump shows up at about 4 T. At 2 K, there are two small jumps at about 5.7 and 6.5 T. These anomalies are associated with the transitions from the CF state to the intermediate ICAF states. For $H \parallel c$ and at 2 K, there are a sharp jump and a weak transition at about 0.9 and 5.2 T, respectively. Since the magnetic moment is about 2.5 $\mu_B$/Co at higher field, this 5.2 T transition is likely the polarization transition. The lower-field jump is due to the transition from the CF state to the ICAF state. At 5 K, the 0.9 T transition evolves into two transitions at about 0.6 and 1.3 T. All these phenomena are consistent with those earlier studies.\cite{Rogado, Szymczak, Yasui, Wilson2, Yen, Fritsch}

\begin{figure}
\includegraphics[clip,width=6.5cm]{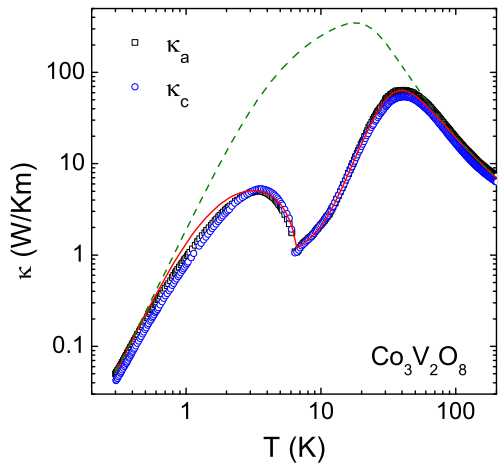}
\caption{(Color online) Temperature dependencies of thermal conductivity of Co$_3$V$_2$O$_8$ single crystals in zero field. The heat current was applied along the $a$ or $c$ axis. The solid line is the fitting to the $\kappa_a$ data using the Debye model with including the phonon scattering effect by the critical fluctuations of magnetic transitions. The dashed line shows the calculated results with switching off the magnetic scattering and displays a standard behavior of phonon heat transport.}
\end{figure}

Figure 2 shows the temperature dependencies of $\kappa_a$ and $\kappa_c$ of Co$_3$V$_2$O$_8$ single crystals in zero field. The temperature dependence of $\kappa$ is complicated but the $\kappa$ are nearly isotropic along the $a$ and the $c$ axis, suggesting a purely phononic transport. The most remarkable feature of the $\kappa(T)$ is a double-peak profile at 3 and 40 K. The positions of these two peaks are rather different from that of phonon peak in insulators, which usually locates at 10--20 K.\cite{Berman} Between two peaks, there is a deep minimum or dip at 6.7 K, which is very close to the transition temperature from the higher-$T$ phase to the CF ground state. This is a common phenomenon in those insulators exhibiting magnetic phase transitions.\cite{Berggold, Wang_HMO, Chen_MCCL, Wang_TMO, Wu_CHC, Buys} Naturally, such kind of dip is caused by strong phonon scattering by the critical spin fluctuations associated with the magnetic phase transition.

The phonon thermal conductivity can be described by a classical Debye model,\cite{Berman, Ziman}
\begin{equation}\label{eq:eps}
\kappa_{ph}=\frac{k_B}{2\pi^2 v_p}\left(\frac{k_B}{\hbar}\right)^3
T^3\int_0^{\Theta_D/T} \frac{x^4e^x}{(e^x-1)^2} \tau(\omega,T)dx,
\end{equation}
in which $x = \hbar\omega/k_BT$, $\omega$ is the phonon frequency, $\Theta_D$ is the Debye temperature, and $\tau(\omega,T)$ is the mean lifetime or scattering rate of phonons. The phonon relaxation is usually defined as
\begin{equation}\label{eq:eps}
\tau^{-1} = v_p/L + A\omega^4 + BT\omega^3\exp(-\Theta_D/bT) +   \tau_{m}^{-1},
\end{equation}
where the four terms represent the phonon scattering by the grain boundary, scattering by the point defects, the phonon-phonon Umklapp scattering, and the magnetic scattering associated with the magnetic phase transitions, respectively. $v_p$ is the phonon velocity, $L$ is the sample width, $\Theta_D$ is the Debye temperature, and $A$, $B$ and $b$ are adjustable parameters. According to Kawasaki's phenomenological theory,\cite{Kawasaki, Rivers} the magnetic scattering rate can be expressed as
\begin{equation}\label{eq:eps}
\tau_{m}^{-1} = C \omega^2 T\left[D(1-T/T_c)^{\alpha}+\omega\right]
\end{equation} for $T > T_c$, and
\begin{equation}\label{eq:eps}
\tau_{m}^{-1} = C \omega^2 T\left[D(T/T_c-1)^{\alpha'}+\omega\right]
\end{equation} for $T < T_c$.
Here, $C$, $D$, $\alpha$, and $\alpha'$ are free parameters. Using formulas (1--4), the $\kappa_a(T)$ data are fitted. Among the parameters of this formula, $L =$ 3.4 $\times$ 10$^{-4}$ m is the sample width and $T_c$ is selected as 6.7 K, the dip position of the $\kappa(T)$ curve. Figure 2 shows the best fitted result with other parameters $v_p =$ 2010 m/s ($\Theta_D =$ 267 K), $A =$ 1.3$\times$10$^{-43}$ s$^3$, $B =$ 1.0$\times$10$^{-30}$ s$^2$/K, $b =$ 2.7, $C =$ 1.5$\times$10$^{-4}$ s$^2$K, $D =$ 2.6$\times$10$^{13}$ s$^{-1}$, $\alpha =$ 2.9, and $\alpha' =$ 1.05. Based on the fitting to the $\kappa(T)$ with Debye model, one can switch off the magnetic scattering by setting $\tau_{m}^{-1} =$ 0. The phonon thermal conductivity obtained in this way is much larger than the experimental data, as shown in Fig. 2. (Similar fitting result can also be obtained for the $\kappa_c$ data.) This analysis indicates that the critical fluctuations associated with the magnetic phase transitions produce strong scattering of phonons. If one can suppress the magnetic scattering, for example, by applying magnetic field, it is possible to increase the $\kappa$ significantly.\cite{Berggold, Wang_HMO, Li_TTO}

\begin{figure}
\includegraphics[clip,width=8.5cm]{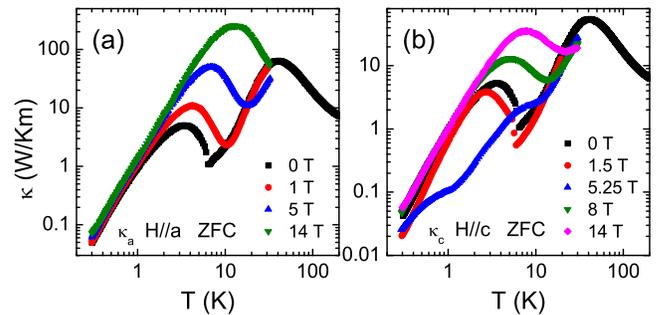}
\caption{(Color online) The $a$-axis and $c$-axis thermal conductivities of Co$_3$V$_2$O$_8$ single crystals versus temperature. Since the magnetic fields were applied along the direction of the heat current, the demagnetization effect is negligibly small for these long-bar shaped samples. The samples were firstly cooled down to the base temperature in zero field (ZFC) and then the magnetic fields were applied. The measurements were carried out with slowly warming up.}
\end{figure}

Figures 3(a) and 3(b) shows the temperature dependencies of $\kappa_a$ with $H \parallel a$ and $\kappa_c$ with $H \parallel c$, respectively. It can be seen that the magnetic field induces drastic changes in the magnitude and the temperature dependence of $\kappa$. In general, the magnetic fields indeed increase the $\kappa$, particularly at the temperature regime between the two peaks. Actually, with increasing field along the $a$ axis, the dip of $\kappa(T)$ at 6.7 K gradually moves to higher temperature, accompanied with the strong enhancement of $\kappa$. It is expectable that in high-field limit ($>$ 14 T), the dip will completely disappear and there would be only one peak locating at $\sim$ 20 K with a magnitude as large as several hundreds of W/Km. It is compatible with the Debye calculation with switching off the magnetic scattering. The $c$-axis magnetic field affects the $\kappa$ in a more complicated way. On one hand, high field along the $c$ axis also induces strong increase of $\kappa$, although it is weaker than that for $H \parallel a$. It is reasonable since the spin polarization is more difficult for $H \parallel c$. On the other hand, lower fields of 1.5 and 5.25 T can suppress the $\kappa$.

\begin{figure}
\includegraphics[clip,width=8.5cm]{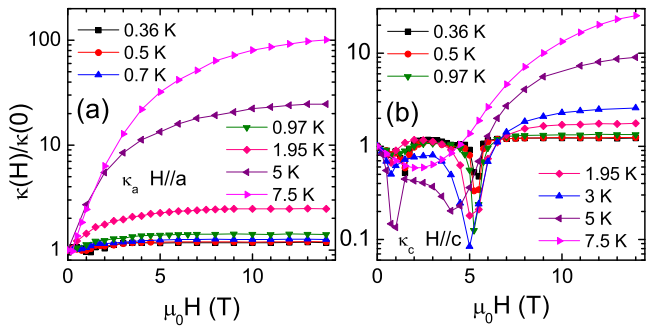}
\caption{(Color online) The $a$-axis and $c$-axis thermal conductivities of Co$_3$V$_2$O$_8$ single crystals versus magnetic field. The data were measured with ascending field step by step after samples zero-field cooled to each temperature. The magnetic fields were applied along the direction of heat currents.}
\end{figure}

The detailed field dependencies of $\kappa$ are displayed in Fig. 4. It is found that for $H \parallel a$, the $\kappa_a$ changes weakly at subkelvin temperatures: as shown in Fig. 4(a), the $\kappa_a$ show a small increase (10--20 \%) with increasing field up to 14 T. This indicates that there are rather weak magnon scattering in the low-$T$ zero-field state. It is also compatible with the conclusion from the zero-field $\kappa(T)$ data. However, at higher temperatures, the $\kappa_a$ shows strong increase with increasing field along the $a$ axis. In particular, at 7.5 K the $\kappa$ continuously increase to a large value at 14 T, which is about 100 times larger than the zero-field value.

The more complicated $\kappa(H)$ behavior for $H \parallel c$ can be clearly seen in Fig. 4(b). At 7.5 K, the $\kappa(H)$ exhibits a broad valley-like feature at several tesla. At lower temperatures, there are two ``dips" in the $\kappa(H)$ curves. The positions of these ``dips" are weakly dependent of the temperature and have good correspondence to the magnetic transitions shown by the magnetization data. It is already known that at low temperatures, the $c$-axis field can induce magnetic transitions from the zero-field CF state to an intermediate ICAF phase and then to the paramagnetic state at about 1 and 5 T, respectively.\cite{Szymczak, Yasui, Yen} It is also a rather common phenomenon in the magnetic materials that the phonons can be strongly scattered by the magnetic excitations populated at the critical fields of various field-induced transitions.\cite{Wang_HMO, Chen_MCCL, Wang_TMO, Wu_CHC, Buys} The strong suppression of $\kappa$ at these two critical fields is also evidenced in the $\kappa(T)$ data in 1.5 and 5.25 T fields, as shown in Fig. 3(b). It is further found that the low-$T$ $\kappa(H)$ for $H \parallel c$ exhibit irreversibility with changing field. (See Fig. S2 of the Supplementary Material.\cite{Supplementary})

It is useful to give a bit more discussions on the phonon scattering effect. First, the above calculation of $\kappa(T)$ data based on Debye model indicates that the magnetic scattering is particularly strong and can be suppressed by magnetic field. It should be noted that the $\kappa$ continuously increases with increasing field up to 14 T (see, for example, the 7.5 K $\kappa(H)$ curve in Fig. 4(a)), while the spins are polarized at around 6 T (7.5 K). There are two possible reasons for this phenomenon: (i) For an AF correlated material, when magnetic field drives the system into a polarized state, the spectrum of magnetic excitations is dispersive and gapped, with the gap increasing with field. Therefore, the magnetic excitations becomes more difficult to be populated in higher field, which results in a continuous suppression of phonon scattering with keep increasing field.\cite{Wang_HMO} (ii) Other less important effects on phonon transport could be related to the magneto-elastic coupling or magnetostriction effect.\cite{Samantaray} Second, the magnetic scattering of phonons is weak at subkelvin temperatures. It can be seen from Debye formula that the phonon scattering rate is dominated by the boundary scattering. The mean free path of phonons, $l$ , at low temperatures can be calculated using the kinetic formula $\kappa_{ph} = Cv_pl/3$,\cite{Berman, Sun_Comment} where $C = \beta T^3$ is the phonon specific heat at low temperatures, $v_p$ is the average velocity. The $\beta$ and $v_p$ values can be obtained from the Debye temperature. Based on the zero-field $\kappa_a$ data, the $l$ at the lowest temperature (0.3 K) is calculated to be 0.18 mm. This is comparable to the sample width ($L$), indicating that the microscopic scattering of phonons is negligibly weak and the so-called boundary scattering limit is nearly achieved at 0.3 K. It is therefore reasonable that at subkelvin temperatures the magnetic field can hardly increase the phonon thermal conductivity.

In summary, the magnetic fields can change the $\kappa$ of Co$_3$V$_2$O$_8$ single crystals in an extremely large scale. In particular, a 14 T field along the $a$ axis can increase the $\kappa$ at 7.5 K up to 100 times. For $H \parallel c$, the magnitude of $\kappa$ can be suppressed down to $\sim$ 8\% at some particular low fields and can be enhanced up to 20 times at 14 T. To our knowledge, such a large magnetothermal effect of phonon heat transport has not been observed in any other magnetic materials.\cite{Berggold, Wang_HMO, Li_TTO} Both the temperature and field dependencies of $\kappa$ demonstrate that the phonons are strongly scattered at the regime of magnetic phase transitions. Magnetic field can suppress the magnetic transition and magnetic fluctuations and significantly recovers the phonon thermal conductivity. One may note that it is a common phenomenon that the phonons can be strongly scattered by the critical fluctuations of magnetic transitions.\cite{Ando, Berggold, Yan, Wang_HMO, Li_TTO, Gofryk} However, in most materials such scattering effect is hardly to change the thermal conductivity in a large scale. Co$_3$V$_2$O$_8$ differs from other materials in the regard that there are five successive magnetic transitions in a narrow temperature range (12--6 K), which produces particularly strong phonon scattering. This suggests that the magnetic materials with complex magnetic structures and magnetic transitions would be able to exhibit similar behavior. It is useful for community to find heat-switch materials working at higher temperatures and in lower magnetic field.

~\\
{\bf SUPPLEMENTARY MATERIAL}

See supplementary material for crystal structure, phase diagram and some low-temperature thermal conductivity data.

\begin{acknowledgements}

This work was supported by the National Natural Science Foundation of China (Grant Nos. 11374277, 11574286, 11404316, U1532147), the National Basic Research Program of China (Grant No. 2015CB921201), and the Opening Project of Wuhan National High Magnetic Field Center (Grant No. 2015KF21).

\end{acknowledgements}

\pagebreak
\widetext
\begin{center}
\textbf{\large Supplementary Material for\\ ``Heat switch effect in an antiferromagnetic insulator Co$_3$V$_2$O$_8$''}
\end{center}

\setcounter{equation}{0}
\setcounter{figure}{0}
\setcounter{table}{0}
\setcounter{page}{1}
\makeatletter
\renewcommand{\theequation}{S\arabic{equation}}
\renewcommand{\thefigure}{S\arabic{figure}}
\renewcommand{\bibnumfmt}[1]{[S#1]}
\renewcommand{\citenumfont}[1]{S#1}

Figure S1 shows the $H-T$ phase diagrams of  Co$_3$V$_2$O$_8$, reported by Ref. \onlinecite{Yen}, and the crystal structure.

It is further found that the low-$T$ $\kappa(H)$ for $H \parallel c$ exhibit irreversibility with changing field. As shown in Fig. S2, the $\kappa_c(H)$ isotherms measured with field increasing are smaller than those with field decreasing, forming a clear hysteresis at low temperatures. The irreversibility weakens with increasing temperature. At 1.95 K, there is almost no hysteresis of $\kappa(H)$. Remember that the magnetization at 2 K is also nearly reversible with changing field. Note that the hysteresis of $\kappa(H)$ of a magnetic material itself is not strange. The irreversibility has been known to often appear at the first-order magnetic transition, such as the spin-Peierls to AF order,\cite{Takeya} the liquid-gas-like transition in spin-ice compounds,\cite{Kolland, Li_DYTO} and the magnetic structure transition in multiferroic DyFeO$_3$,\cite{Zhao_DFO} etc. The $\kappa(H)$ hysteresis of Co$_3$V$_2$O$_8$ indicate that the low-field magnetic transition is a first-order one, in contrast to the continuous polarization transition at 5 T.

It should be noted that at subkelvin temperatures there are actually two transitions in low fields. At 0.36 K, the two transitions are well separated in the field ascending curve and are locating at 0.75 and 1.5 T. It can also be seen that the irreversibility of $\kappa(H)$ is much more significant at the second transition. With increasing temperature to 0.97 K, the position of the first transitions does not change while the second one moves to 1.25 T, with the weakening of the irreversibility. At even higher temperature of 1.95 K, the second transition seems to completely disappear or merge with the first one at 0.75 T. In this regard, it is not strange that the magnetization curve at 2 K displays only one transition at $\sim$ 0.9 T, which is nearly reversible. Meanwhile, the irreversibility of $\kappa(H)$ disappears. Another weaker but notable feature of data shown in Fig. S2 is that there is a small irreversibility of $\kappa(H)$ at about 4 T. This reversibility is clearly discernable at 0.36 and 0.5 K and disappears at higher temperatures. At 0.36 and 0.5 K, the $\kappa(H)$ show a weak discontinuity when increasing field across 4 T, which seems to be magnetic transition. These findings indicate that the phase diagram of Co$_3$V$_2$O$_8$ with $H \parallel c$ is more complicated than what has been known. As discussed above, the 0.9 T transition at $T \approx$ 2 K was found to be a transition from the low-field CF state to the CAF state (with $\delta =$ 1). The present result at ultra-low temperatures reveals that there are likely two first-order transitions at 1.25 and 4 T, which are located in the CAF ($\delta =$ 1) phase of original phase diagram. At present, the natures of these two transitions are not known and call for the neutron scattering measurements at subkelvin temperatures.

\begin{figure}
\includegraphics[clip,width=8cm]{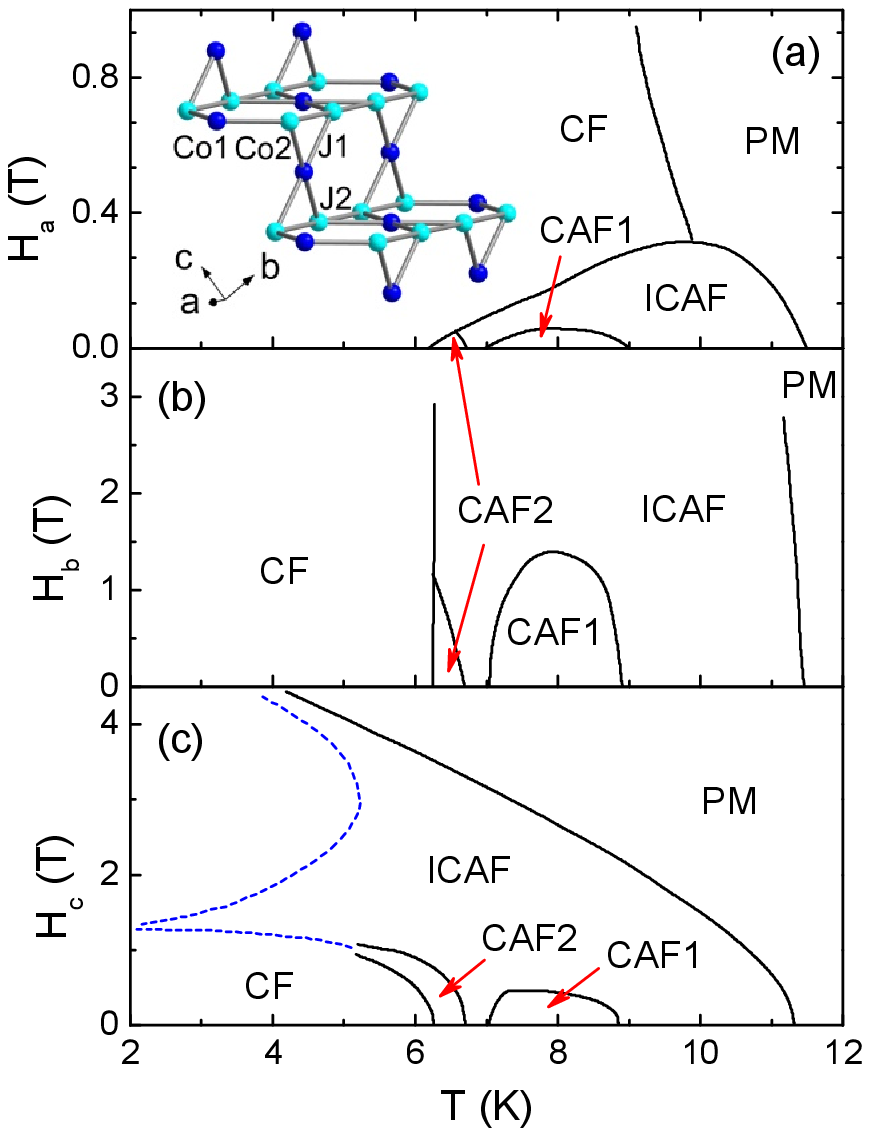}
\caption{(Color online) Schematic plots of the $H-T$ phase diagrams of  Co$_3$V$_2$O$_8$, reported by Ref. \onlinecite{Yen}. The magnetic field are along the $a$, $b$ or $c$ axis. Five phases have been determined: paramagnetic phase (PM), incommensurate antiferromagnetic phase (ICAF), the first commensurate phase with $\delta =$ 1/2 (CAF1), the second commensurate phase with $\delta =$ 1/3 (CAF2); the weak ferromagnetic phase (CF). The inset shows the crystal structure. Two crystallographically inequivalent Co$^{2+}$ sites exist, that is, the cross-tie sites Co(1) forming the apices of the isosceles triangles and the spine sites Co(2) forming the bases of the triangles. The structure consists of the Co(2)$-$Co(1)$-$Co(2) isosceles triangles and the ratio of the Co(2)$-$Co(2) and Co(1)$-$Co(2) bond distances is about 1.003.\cite{Rogado, Sauerbrei}}
\end{figure}

\begin{figure}
\includegraphics[clip,width=8.5cm]{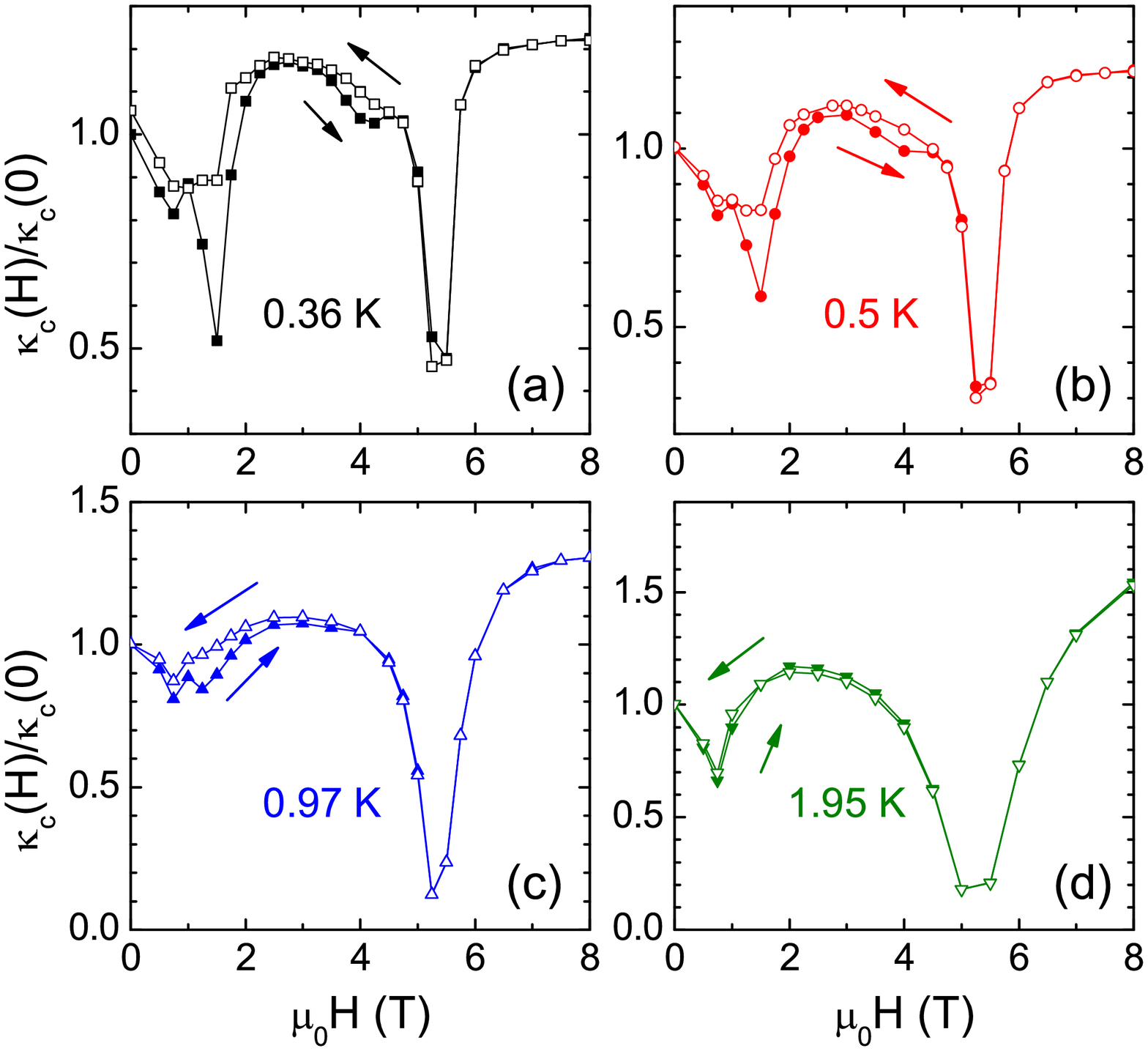}
\caption{(Color online) Irreversibility of the low-temperature $\kappa_c(H)$ isotherms of Co$_3$V$_2$O$_8$ single crystals with $H \parallel c$. The data shown with solid symbols were measured in the ascending field after the sample is cooled in zero field, while the open symbols show the data with the descending field, as indicated by arrows.}
\end{figure}

\end{document}